\documentclass[11pt]{JHEP3}
\usepackage[T1]{fontenc}

\usepackage{latexsym}
\usepackage[latin1]{inputenc}
\usepackage{amsmath,epsfig}
\usepackage{amssymb,amsfonts}
\usepackage{bm}
\usepackage{cite}
\usepackage{mathrsfs}
\usepackage{cancel,empheq,dsfont,overpic,subcaption,subeqnarray}
\usepackage[normalem]{ulem}
\usepackage{slashed, braket}
\usepackage{xcolor}

\usepackage{graphicx}
\usepackage{float, soul, parskip, lmodern}
\usepackage{caption}
\usepackage{subcaption}
\usepackage{gensymb}
\usepackage{bbold}
\usepackage{amsmath}
\usepackage{longtable}
\usepackage{color}
\usepackage{bbm}
\usepackage{tikz}
\usepackage[compat=1.1.0]{tikz-feynman}
\usepackage{comment}

\def\be{\begin{equation}}
\def\ee{\end{equation}}

\newcommand{\bear}{\begin{eqnarray}}
\newcommand{\bea}{\begin{eqnarray}}
\newcommand{\eear}{\end{eqnarray}}
\newcommand{\eea}{\end{eqnarray}}
\newcommand{\ba}{\begin{array}}
\newcommand{\ea}{\end{array}}
\newcommand{\bi}{\begin{itemize}}
\newcommand{\ei}{\end{itemize}}

\def\a{\alpha}		\def\b{\beta}				
\def\e{\varepsilon}						
		\def\k{\kappa}		\def\l{\lambda}		\def\m{\mu}
\def\n{\nu}			\def\x{\xi}

		\def\D{\Delta}

\renewcommand{\b}[1]{\textbf{#1}}

\def\II{\relax{\rm I\kern-.18em I}}

\def\e{\epsilon}
\def\m{\mu}
\def\n{\nu}

\def\sp{\;\;\;,\;\;\;}

\def\a{\alpha}
\def\b{\beta}

\def\l{\lambda}

\def\<{\big\langle}
\def\>{\big\rangle}

\def\nn{\nonumber}

\def\text#1{{\rm #1}}

\def\hri#1#2{\href{http://arxiv.org/abs/#1}{[ArXiv:#1]#2}}
\def\hre#1#2{\href{http://arxiv.org/abs/#1/#2}{[ArXiv:#1/#2]}}
\def\hrj#1#2{\href{https://doi.org/#1}{#2}}

\title{Emergent neutrinos from heavy messengers}

\author{\large Pascal~Anastasopoulos$^{1}$\footnote{pascal.anastasopoulos@univie.ac.at }, Elias~Kiritsis$^{2}$\\
 ~\\
{$^1$ \href{https://mathphys.univie.ac.at}{Mathematical Physics Group, Department of Physics, University of Vienna}, \\
Boltzmanngasse 5, 1090 Vienna, Austria}\\
 ~\\
{$^2$ \href{http://hep.physics.uoc.gr}{Crete Center for Theoretical Physics}, Institute for Theoretical and Computational Physics,
Department of Physics,
University of Crete, 70013, Heraklion, Greece}\\
  \centerline{and}\\
\href{http://www.apc.univ-paris7.fr}{Universite de Paris, CNRS, Astroparticule et Cosmologie,  F-75006 Paris, France}\\
}

\preprint{
UWThPh 2021-26\\ 
CCTP-2022-1\\
ITCP-IPP 2022/1
}

\abstract{Fermionic bound states (mesinos) of messengers can play the role of right-handed neutrinos and due to their heavy masses they can realize the seesaw mechanism providing light and heavy sterile neutrinos. We study simple models with a single mesino to present the proof of principle. We extend our analysis to three mesino states where after the seesaw mechanism,  three light and several heavy sterile neutrinos are obtained.

}


\begin{document}
\maketitle
\flushbottom


\section{Introduction and conclusions}
\label{sec:intro}

Neutrino physics is one of the most exciting topics of high-energy physic.
Experimental bounds show that neutrinos are massive but much lighter than any other known massive particle in the SM,
requesting physics beyond the SM (see \cite{Deppisch:2015qwa} and citations within).
In this framework, various scenarios have been developed where new particles couple to the gauge-invariant fermionic portal $LH$ of the SM and trigger the seesaw mechanism (type I \cite{Minkowski:1977sc, Mohapatra:1979ia,Yanagida:1979as,Gell-Mann:1979vob, Schechter:1980gr}, type II \cite{Magg:1980ut,Cheng:1980qt, Lazarides:1980nt, Mohapatra:1980yp}, type III \cite{Foot:1988aq,Dev:2012sg, BhupalDev:2012jvh}) that predict the light masses of the neutrinos.
They also provide exciting phenomenology from low-energy accelerator experiments to astroparticle physics and cosmology.


In the most popular scenarios, the SM is extended by extra fermions, the $right$-$handed$ $neutrinos$ $N_R^i$ that trigger the type I seesaw mechanism by a large Majorana mass term. This $N_R^i$ can be a fundamental or a composite particle that emerges from a hidden sector of the theory that only couples to the SM via messengers.

Focusing on composite $N_R$, including a hidden sector, we can have two different classes of models.
\begin{enumerate}

\item $N_R$ as fermionic operators of a hidden sector.

In that case, the right-handed neutrinos are fermionic gauge-invariant operators of the hidden sector.
In this case the right handed neutrinos are  like the baryons baryons in QCD.
Several works in the past have already discussed this possibility \cite{Arkani-Hamed:1998wff, Okui:2004xn, Grossman:2008xb, Grossman:2010iq, Hundi:2011et, Robinson:2012wu, Biondini:2017fut, Davoudiasl:2017zws} focusing mainly on the type I (canonical) seesaw mechanism \cite{Minkowski:1977sc, Mohapatra:1979ia,Yanagida:1979as,Gell-Mann:1979vob, Schechter:1980gr}. 

\item $N_R$ as fermionic operators of messenger fields.

In this case, the right-handed neutrinos are bound states of, for example, a fermionic and a bosonic messenger field. Both will be charged under the SM and the hidden gauge group, forming a meson-like combination \footnote{More complicated combinations of messengers are also allowed, but we keep our discussion to the simplest case.}.

The key point is that these bound states are massive with masses of the order of the messenger scale, and for very heavy messengers, they can naturally trigger the seesaw mechanism providing light and heavy neutrinos.

This case has not been studied in the past, and it is the subject of this article.

\end{enumerate}

In the present work, we explore the possibility where right-handed neutrinos are emergent fields in a holographic-inspired scenario where the complete theory is described bya  four-dimensional QFT and consists of three sectors \cite{K},
a) the SM sector, which contains all known particles.
b) A hidden sector that does not directly couple to the SM. This sector is UV-complete, and for simplicity, we will assume a large $N$-strongly coupled gauge theory.
c) The messenger sector comprises massive bi-fundamental fields charged under the SM and the hidden gauge group. Their masses $M$ are above all other scales in this framework.

The heavy messengers are integrated out at energies below the messenger scale, $M$. We obtain an effective action for the low-energy theory that involves the SM and the hidden theory coupled with double (or multiple) trace interactions, all irrelevant \footnote{There are two exceptions, where couplings of the SM with the hidden sector might be relevant:
a) The Higgs mass term. This term is related to the hierarchy problem, which can be avoided if all operators of the hidden theory have dimensions well above two.
That is the situation which we will assume. 
b) The field strength of the hypercharge. This mixing~has~been~studied~in~a similar context in \cite{gravi}.
}.
Therefore, composites  of the hidden sector behave like weakly-coupled particles coupled  to the SM, and those protected by symmetries remain light, providing interesting phenomenology \cite{K, Anastasopoulos:2021osp}. It is worth mentioning the emergent ``graviton'' \cite{tensor}, the emergent axions \cite{axions}, the emergent vector fields \cite{vector} graviphotons \cite{gravi} and a new portal for dark matter \cite{Anastasopoulos:2020gbu}. All these fields have a compositeness scale below which they behave like fundamental particles, and above, they enter in a non-local, non-standard description.

In this holographic inspired framework, $N_R$ neutrinos can either be operators of the hidden sector (baryonic type) or operators of the messenger sector (mesonic type). Our goal is to explore the second possibility.

\subsection*{Results and outlook}

This paper explores the possibility that the right-handed neutrinos are realized as bound states of heavy messenger fields.
The framework is described in \cite{K}, and it consists of three sectors (SM, hidden, and messenger sectors).
\bi
\item First, we parametrize the effective action of the heavy messenger bound states.
Based on effective field theory principles, we provide the couplings of these mesons/mesinos among themselves and the SM operators.
Being heavy, these fields usually decouple from the low energy action, however, they can trigger a seesaw mechanism and provide light and heavy states. Results are presented in section \ref{sec:messenger bound-states}.

\item To present the ``proof of principle'', we analyze the simplest toy-models with a single bosonic and a single fermionic messenger that provide heavy mesinos that couple to the $LH$ fermionc operator of the SM. We present two cases where mesinos are Dirac or Majorana fermions triggering the seesaw mechanism differently. The result is always a light and some heavy sterile neutrinos. Results can be found in section \ref{SOMEexamples}.

\item Next, we generalize our toy models to obtain semi-realistic models with three light SM neutrinos. We assume a single fermionic and three bosonic messengers, copies of the examples analyzed in section \ref{SOMEexamples}, and we show that the three mesinos obtained from these messengers can play the role of the right-handed neutrinos leading to three SM neutrinos and some sterile neutrinos. Analysis and results are presented in section \ref{3 Neutrino Cases}.

\item Composite right-handed neutrinos is an exciting possibility providing different options. We list some of them below.

\bi

\item First of all, we can assume three different/independent hidden sectors with any of the $SU(N_i)$, $SO(2N_i+1)$, $Sp(N_i)$ with large $N_i$ and each of them to connect with the SM via one bosonic and one fermionic messenger (three copies of the toy-model presented in section \ref{SOMEexamples}). That would generate three copies of mesons/mesinos. The messengers can have different mass scales generating a hierarchy between the masses of the resulting neutrinos.

\item Another option is the three right-handed neutrinos to be produced by combinations of mesonic (the mechanism presented here) and baryonic operators coming entirely from the hidden sector \cite{Arkani-Hamed:1998wff, Okui:2004xn, Grossman:2008xb, Grossman:2010iq, Hundi:2011et, Robinson:2012wu, Biondini:2017fut, Davoudiasl:2017zws}.

For example, we can build models with two mesonic right-handed neutrinos (with different messenger masses, as described above) and one baryonic type of neutrino. That will trigger the seesaw mechanism with at least three scales.

\ei

\item One issue of these models, and all models with a messenger sector, is the hierarchy problem. Mesonic operators of the messenger sector couple to the Higgs square terms in the action and can lift the vevs of the Higgs in undesired values. Eventually, we have to fine-tune various parameters of these models to bypass this issue.
\ei

In conclusion, we show that in holography-inspired models, right-handed neutrinos coming from the messenger sector is an interesting possibility with many different options, and deeper analysis is required at the theoretical and phenomenological level.

\section{Messenger bound-states}
\label{sec:messenger bound-states}

This section gives more details about our framework, and we parametrize the effective action of bound states of messenger fields.

Consider an UV-complete, 4D quantum field theory that consists of three sectors.
\bi
\item[a)] The SM sector contains all known fields (quarks, leptons, gauge fields, Higgs).
At first, we consider the minimal content of the  SM, and all fields beyond them will in the hidden sectors. Later, we will loosen the standard definition and investigate minimal extensions of the SM.

\item[b)] The combined theory including the hidden sectors  is considered to be UV-complete. Therefore, all couplings of the theory at the extreme UV region are renormalizable.
The rank of the hidden sector  gauge group is assumed to be large. We do not for the moment impose other restrictions on its spectrum apart from the fact that there is a regular t' Hooft large N limit.
 However, we can assume $SU(N_i)$ factors with $N_i$ very large to simplify our discussion. Generically the hidden QFT will be assumed strongly coupled.
Therefore, the hidden theory is a generalized large-N gauge theory with scalars and  fermions beyound the gauge bosons. They transform in various representations of the hidden gauge group (fundamentals, adjoints, and
two-index symmetric and antisymmetric representations).

\item[c)] The messenger sector contains fields that couple the hidden theory and the SM at some high energy scale. They are generically are bifundamentals under the hidden gauge group and the SM gauge group.
  The messenger fields will be assumed to have large masses, of order ${M}$, much above all the other scales of our framework.
\ei
As we have already mentioned, we will mainly focus on the messenger sector. In general, we have
\bea
\phi_{ai} ~:~~\textrm{bosonic messengers} ~,~~~~~~\psi_{ai}^\a ~:~~\textrm{fermionic messengers}
\label{generic messengers}\eea
They are bifundamental fields, charged under some SM gauge groups, denoted by the indexes $i,j$ and charged under some hidden index $a,b$ from the hidden gauge group. The $\a,\b$ are the space-time SU(2) spinor indices. Notice also that fermionic messengers are non-chiral.

These generic messengers \eqref{generic messengers} give rise to three bilinears that are gauge-invariant under the hidden gauge group.
Because of the strong hidden color interaction, such bilinears will give rise to tightly bound mesons in the theory with masses that are large, but will be lower that the sum of the masses of theirr constituents due to the strong attractive interaction.\footnote{A weakly coupled bound state of two ``quarks" of mass $M$ has approximate mass $2M$. But at very strong coupling in a scale invariant theory as N=4 sYM, the mass scales as $M/\lambda$
where $\l$ is the 't Hooft coupling, and therefore it can be hierarchically smaller than $M$ is the coupling is large.}
They are given as\footnote{$M$ is setting up the messenger masses. However, if the UV CFT is not free, then the notion of messenger mass has to be defined indirectly. It is the relevant coupling multiplying the linear combinations of ``meson'' operators $\Phi_{i\bar j}$, $X_{\bar ij}$, that are SM singlets.}
\be
\Psi^{\a}_{i\bar j}=\psi^{\a}_{ai}\phi^*_{aj}\sp \Phi_{i\bar j}=\phi_{ai}\phi^*_{aj}\sp X_{\bar ij}=\bar \psi^{\a}_{ia} \psi^{\b}_{aj}\e_{\a\b}
\label{a1}\ee
$X_{i\bar j}$, $\Phi_{ij}$ are complex bosonic operators while $\Psi^{\a}_{ij}$ are fermionic.

The UV action, including the mass terms, is
\be
S=S_{UV}+\int d^4 x \delta_{ij}\left(M^2\Phi_{ij}+MX_{ij}\right)
\label{a2}\ee
where $S_{UV}$ is the action of the UV CFT, and for simplicity\footnote{As long as we are interested in determining the scale dependence of the effective action, this is without loss of generality.} we chose the same masses for all messengers.

For a general UV CFT that is not free, the dimensions of the mesonic operators will depart from their free field values, which we denote by $\Delta_{X,\Phi,\Psi}$ respectively.
From unitarity they must satisfy
\be
\Delta_{\Psi}\geq {3\over 2}\sp \Delta_{\Phi,X}\geq 1
\label{a3}\ee
Now, the mass terms are generalized as follows
\be
S=S_{UV}+\int d^4 x \delta_{ij}\left(M^{4-\Delta_{\Phi}}\Phi_{ij}+M^{4-\Delta_{X}}X_{ij}\right)
\label{a4}\ee
and this is what we will consider from now on.

There is another scale in the hidden theory that we call $\Lambda$ and is the analog of $\Lambda_{QCD}$. It is the coupling associated to the relevant operator $O_{\Lambda}$ of UV scaling dimension $\Delta_{\Lambda}$. It controls a relevant interaction that drives the hidden gauge theory to the IR and may be much larger or smaller than the associated SM scales. We may therefore write
\be
S=S_{UV}+\int d^4 x \left[\Lambda^{4-\Delta_{\Lambda}}O_{\Lambda}+ \delta_{ij}\left(M^{4-\Delta_{\Phi}}\Phi_{ij}+M^{4-\Delta_{X}}X_{ij}\right)\right]
\label{a4aa}\ee
for the relevant part of the action that affects the behavior of the mesonic operators.


\subsection{From two-point functions to effective actions}

In appendix \ref{a-2p}, we show how the two-point functions of the composite fields $X_{i\bar j},\Phi_{i\bar j},\Psi_{i\bar j}$ depend on $M,\Lambda$, and we extract the non-derivative non-linear interactions. Our results, for $M\gg \Lambda$, $p^2\ll M^2$, are listed below.
\bi
\item For bosonic operators $\chi_{i\bar j}= \Phi_{i\bar j}, X_{i\bar j}$ we obtain from \eqref{a27a}
\be
\langle \chi_{i\bar j}\chi_{k\bar l}\rangle\sim \delta_{i,\bar l}\delta_{\bar j,k}M^{2\Delta_{\chi}-4}\left[1-\kappa_\chi{p^2\over M^2}+\cdots\right]\sp M\gg \Lambda\sp p^2\ll M^2
\label{27}\ee
Assuming $M\gg \Lambda$, we can evaluate the action including non-derivative non-linear interactions \eqref{a33} (see appendix \ref{a-2p}),
\be
S_{\Phi,X}=
 -\int d^4x \left(-{M^2\over \k_\chi}+{\square}\right) {\cal X}(x)^2
 +\sum_{n=3}^{\infty}g_n{M^{4-n}\over \k_\chi^{n/ 2}} {\cal X}(x)^n
\ee
where $\k_\chi$ is a constant, generically of order 1 and $g_n$ are dimensionless couplings.
Also, ${\cal X}$ are canonically normalised fields, defined from $\chi$ with the relation \eqref{a31}
\bea
{\cal X}\equiv
{\sqrt{\k_\chi}\over M^{\Delta_\chi-1}}\chi(x)-{4c_M\sqrt{\k_\chi}\over 4-\Delta_\chi}{M}\eea
Notice that the UV vev of the mesons $\chi$, proportional to the large messenger scale, disappear from our action.

\item The $X'_{i\bar j}$, the non-singlet piece of $X_{i\bar j}$

The singlet piece $X\equiv \delta^{i\bar j}X_{i\bar j}$ has a similar two-point function to (\ref{27}).
The non-singlet piece $X'_{i\bar j}=X_{i\bar j}-{\delta_{i\bar j}\over N}X$ on the other hand has the following two-point function \eqref{a27b}
\be
\langle X'_{i\bar j}X'_{k\bar l}\rangle\sim \left(\delta_{i,\bar l}\delta_{\bar j,k}-{\delta_{i\bar j}\delta_{k\bar l}\over N}\right)M^{2\Delta_{X'}-4}\left[1-\k_{X'}{p^2\over M\Lambda}\cdots\right]\sp M\gg \Lambda\sp p^2\ll M^2
\ee
We rescale to define canonically normalized fields as in \eqref{a34}
\bea
{\cal X}'\equiv {\sqrt{\k_{X'}} \over M^{\Delta_{X'}-{3\over 2} }\Lambda^{1\over 2}}{X'} -{4c_M\sqrt{\k_{X'}}\over 4-\Delta_{X'}}\left({M^3\over \Lambda}\right)^{1\over 2}
\eea
so that the non-derivative non-linear interactions we get \eqref{a36}
\bea
S_{X'}= -\int d^4p \left(-{M\Lambda\over \k_{X'}}+{\square}\right){{\cal X}'(x)^2}+\sum_{n=3}^{\infty}g_n{M^{4-n}\over \k_{X'}^{n/ 2}} {\cal X}'(x)^n
\eea

\item For the mesinos $\Psi^{\a}_{i\bar j}$ we obtain \eqref{a28}
\be
\langle \Psi^{\a}_{i\bar j}\Psi^{\b}_{k\bar l}\rangle\simeq
\delta_{i,\bar l}\delta_{\bar j,k}M^{2\Delta_{\Psi}-4}\left[1-\kappa_\Psi{\slashed{p}^{\a\b}\over M}+\cdots\right]\sp M\gg \Lambda\sp p^2\ll M^2\ee
Therefore, the action for the mesinos (kinetic term), after the rescaling \eqref{a38}
\bea
{\widetilde \Psi}= \sqrt {\k_\Psi} M^{{3\over 2}-\D_\Psi} \Psi
\eea
becomes \eqref{a39} that we reproduce here,
\bea
S_{\widetilde\Psi}&=& -\int d^4x \bar {\widetilde\Psi} \left( i \slashed \partial-{M\over \k_\Psi} \right)\widetilde \Psi + \sum_{n=2}^\infty g_n M^{4-3n} (\bar {\widetilde\Psi} \widetilde \Psi)^n
 \eea

\ei
Note that the normalized composite fields have the same dimensions as the free fields, as long as we remain in the $M\gg \Lambda$ and $M\gg p^2$ range.

\subsection{Interactions}\label{Interactions}

Next, we present all interactions between mesons/mesinos and the SM fields.
\bi
\item Cross terms between the composite fields.
The interaction terms between mesons/mesinos are generically of the form
\bea
S_\textrm{cross-terms} &=& - \int d^4x
\sum_{a+b+c>0}
q_{abc} M^{4-a-b-3c }
{\cal X}(x)^a
{\cal X}'(x)^b
(\bar {\widetilde\Psi} \widetilde \Psi)^c
\label{cross-terms}\eea
where we can have also gamma matrices and derivatives.

\item Interactions between composite fields and the Standard Model fields

Interactions with the SM fields are originating in the mixings of the mesons/mesinos with any gauge invariant operator of the SM.
\bea
S_\textrm{cross-terms} &=& - \int d^4x
\sum_{a,b,c,d,e...}
q_{abc} M^{4-a-b-3c-{3\over 2}d-\Delta} \\
&&~~~~~~~~~~~~~~~~~~~~~~
{\cal X}(x)^a
{\cal X}'(x)^b
(\bar {\widetilde\Psi} \widetilde \Psi)^c (\widetilde \Psi_\a)^d \times
{\cal O}_\Delta[\textrm{SM~fields}] \nn
\label{SMmesinosMIXED}\eea
In appendix \ref{Scale for mixed terms} we derive the scaling of these terms. The leading scale is the messenger scale $M$, and SM scales appear at sub-leading levels.
\ei

\section{Models with emergent neutrinos}\label{SOMEexamples}

In this section, we apply the generic properties analyzed above to simple models focusing on mesinos which can play the role of right-handed neutrinos.
We discuss two different cases, depending on whether mesinos from the hidden sector are complex (Dirac) or real (Majorana) fermions.

\subsection{Messengers in complex representations}

The standard framework mentioned before consists of the SM, hidden, and messenger sectors.
The hidden sector is a large-N $SU(N)$ generalized gauge theory, and the messengers transform under the $SU(N)$ and the SM gauge groups.
In the simplest case, we consider only two messengers transforming in a bifundamental representation of the SM with the hidden sector like
\bea
\ba{llllccccccccclcccc}
 & & & &SU(N) && SU(3) &&SU(2) &&~~~Y~~~& & ~~~& \\
& & \phi_{N,q}~:~~~& (&\Box_N&;& \cdot &,&\cdot &,&q&)&\\
& & \psi_{N,q}~:~~~& (&\Box_N&;& \cdot &,&\cdot &,&q&)&
\ea
\label{messengersCHARGEDmodel1}
\eea
and their charge-conjugates. By $``\cdot"$ we denote the singlets under the $SU(n)$ groups of the SM and the hidden sector. The mesons and mesinos out of the messengers above are singlets under the hidden and SM gauge groups.
\bea
\ba{llllllllllll}
\Phi_1 ~=~\Phi &\sim & \phi_{N,q} \phi^*_{N ,q} ~&
&~~ \textrm{boson}\\
\Psi_1 ~=~\Psi &\sim & \phi_{N,q}^*\psi_{N ,q} ~&
&~~ \textrm{fermion}\\
\Psi_2 ~=~\Psi^c &\sim & \phi_{N,q} \psi_{N ,q}^c ~&
&~~ \textrm{fermion}\\
\Phi_2 ~=~ X &\sim & \psi_{N,q}\bar \psi_{N ,q} ~&
&~~ \textrm{boson}
\ea
\label{MesonsMesionosMODEL1}\eea
This case \eqref{MesonsMesionosMODEL1} is realized in a semi-realistic D-brane model, where the hypercharge is a linear combination of various abelian factors emerging from the different stacks of D-branes \cite{AKT, Ib, rev, ADKS}. In that case, the SM fields, as well as the messengers, are charged under additional abelian symmetries, which are typically anomalous. The anomalies are canceled by axions and generalized Chern-Simons terms rendering the anomalous fields massive \cite{AKR, muon, AL, Bachas, anasta, IQ, irges, Anastasopoulos:2008jt, AbelGoodsell, bianchi}.
Such a model is described in appendix \ref{D-brane SM}.

\subsubsection{The action in the low energy  regime}

Using the generic form of the terms discussed in section \ref{Interactions}, we can present the low energy interaction terms of the mesons/mesinos between themselves and with the SM operators.

The lowest interaction terms for the mesons and mesinos are given by \eqref{cross-terms},
\bea
S_{\text{Meson}} &=&\int d^4x\;\bigg(
M \sum_{i,j=1}^2 q_{ij} \bar \Psi_i \Psi_j + M^{-2} \sum_{i,j,k,l=1}^2q_{ijkl}(\bar \Psi_i \Psi_j)(\bar \Psi_k \Psi_l) ~~~~~~~~~
\label{couplingsONLYMESONS}\\
&&~~~~~~~~~~~+ M^2 \sum_{i,j=1}^{2} g_{ij} \Phi _{i} \Phi _{j} + M \sum_{i,j,k=1}^{2} g_{ijk} \Phi _{i} \Phi _{j} \Phi _{k} + \sum_{i,j,k,l=1}^{2} g_{ijkl} \Phi _{i} \Phi _{j} \Phi _{k} \Phi _{l} \nn\\
&&~~~~~~~~~~~+ \sum_{i=1}^{2} h_{ijk} \Phi _{i} \bar \Psi_j \Psi_k
+ M^{-1}\sum_{i,j=1}^{2} h_{ijkl} \Phi _{i} \Phi _{j} \bar \Psi_k \Psi_l + ....\bigg) \nn
\eea
Notice that, by summing over $\bar\Psi_i\Psi_j$, we include all four bilinears $\bar \Psi \Psi$, $\bar \Psi^c \Psi^c$, $\bar \Psi \Psi^c$, $\bar \Psi^c \Psi$ in \eqref{couplingsONLYMESONS}. If mesinos are charged under any of the SM gauge groups, the last two combinations should be omitted. If, however, mesinos are charged under some broken (gauge) symmetry, these terms are generically present, but their coupling will be suppressed by $\tilde v/M$ where $\tilde v$ is the energy scale of the broken symmetry.

The interaction terms between mesons/mesinos and the SM fields are given by \eqref{SMmesinosMIXED},
\bea
S_{\text{Int}} &=&\int d^4x\; \bigg( M \sum_{i,a=1}^{2} g_{ai} |H_a|^2 \Phi_i
	+ \sum_{i=1}^{2} g_{LH\Psi_i} \bar L_L \Psi_i H^*_2
\label{SMandMesonsMesino}\\
	&& ~~~~~~~~~~ + M^{-1} \bigg\{ \sum_{i=1}^{2}\Phi_{i} \left(H_1^* \bar L_L l_R + H_1^* \bar Q_L d_R + H_2^* \bar Q_L u_R \right) 
	+\sum_{i=1}^{2} g_{12l\Psi_i}H_2 H_1^* \Psi_i\; l_R \nn\\
	&& ~~~~~~~~~~~~~~~~~~~~~~ + \sum_{i=1}^{2} g_{aij} |H_a|^2 \bar \Psi_i \Psi_j
+ \sum_{i,j=1}^{2} g_{2L\Psi_i\Phi_j}H^{*}_2 \bar{L}_L \Psi_i \Phi_{j} \bigg\}+... \bigg)\nn
\eea
A combination of mesinos $\Psi$ and $\Psi^c$ can play the role of the right-handed neutrino.

\subsubsection{Seesaw mechanism with complex mesinos from the messenger sector}

Next, we focus on mass-generating terms. Since mesinos in \eqref{MesonsMesionosMODEL1} are chargeless/singlets we collect
\bea
S_{\text{mass}} =\int d^4x \bigg( g_{\tiny LH\Psi_1} {v_2} ~ \overline{{\n}}_L \Psi + g_{LH\Psi_2} {v_2} ~ \overline{{\n}}_L \Psi^c
+ M_D \bar \Psi\Psi + M_M \bar \Psi^c\Psi+ h.c.\bigg)~~~~~
\label{ActionMASS}\eea
where we absorb all couplings and parameters in the definition of $M_D,M_M$ which are of the same order $\sim M$, and $u_i=\langle H_i\rangle$ the Higgs vevs.
The complete mass term is now $3\times 3$, $(\n^c_L,\Psi,\Psi^c)$.
\bea
{\cal M} =
\left(\ba{cccc}
0& v & v' \\
v& M_M & M_D \\
v'& M_D & M_M
\ea \right)
\eea
with
$$v= g_{LH\Psi_1} {v_2}\sp v'= g_{LH\Psi_2} {v_2}\;.$$
The mass eigenvalues of this mass matrix are
\bea
&&m_{\n_1} \sim v/M_D \nn\\
&&m_{N_1} \sim M_D+M_M \nn\\
&&m_{N_2} \sim M_D-M_M \label{masses1light2heavy}
\eea
and assuming that the difference $v, v' \ll M_D-M_M$, we have a very light neutrino and two heavy neutrinos.

It is also interesting to discuss the case where $\Psi$ is charged under a broken symmetry. That symmetry can be broken either by some Higgs mechanism or by instanton effects (if superficially anomalous). The last case often appears in D-brane realizations of the SM, like the case discussed in the appendix \ref{D-brane SM}, where mesinos are charged under anomalous $U(1)$ symmetries \eqref{MesonsMesionosMODEL1string}.
In this case, Majorana mass terms $\bar \Psi^c \Psi+h.c$ appear after symmetry breaking and their scale $M_M$ is of the order of the symmetry breaking scale and spans from small values up to the messenger scale $M$ (which is assumed to be the largest scale in this framework).
If $M_M$ is of the order of $M\sim M_D$ we have again the mass eigenvalues found in \eqref{masses1light2heavy}.
However, if $M_M\ll M_D$ we have one light and two heavy neutrinos with masses
\bea
m_{\n_1} &\sim& v/M_D \nn\\
m_{N_1},m_{N_2} &\sim& M_D \label{massesBROKENsymmetry}
\eea
since $M_M\ll M_D\sim M$ and the messenger scale dominates the masses of the heavy neutrinos.

\subsection{Messengers in real representations}

In the above case, mesinos transform in complex representations. However, it is interesting to consider the possibility to transform in real representations since that would lead to Majorana fermions triggering the typical type I seesaw mechanism.

As an example, let us assume an $SO(2N+1)$, instead of the $SU(N)$, gauge group for the hidden sector\footnote{Similar results are expected using $Sp(N)$ group for the hidden sector.}. Consider again a fermionic and a bosonic messenger, transforming like the vector representation $V_{2N+1}$ under the orthogonal group and the pseudo-real $\Box_2$ of the $SU(2)$ like
\bea
\ba{llllccccccccclcccc}
 & & & &SO(2N+1) && SU(3) &&SU(2) &&~~~Y~~~& & ~~~& \\
& & \phi~:~~~& (&V_{2N+1}&;& \cdot &,&\Box_2&,&0&)&\\
& & \psi~:~~~& (&V_{2N+1}&;& \cdot &,&\Box_2&,&0&)&
\ea
\label{messengersCHARGEDmodel2}
\eea
These messengers leads to real mesons/mesinos
\bea
\ba{llllllllllll}
\Phi &\sim & \phi \phi ~&
&~~ \textrm{boson}\\
\Psi &\sim & \phi\psi ~&
&~~ \textrm{fermion}\\
X &\sim & \bar \psi \psi ~&
&~~ \textrm{boson}
\ea
\label{MesonsMesionosMODEL2}\eea
Thus, $\Psi$ is Majorana mesino leading to the typical type I seesaw mechanism, as we argue below.

We should stretch here that we could have messengers transforming in real representations even if the hidden sector is an $SU(N)$ group. In this case, the bosonic and fermionic $exceptional$ $messengers$ \cite{K} should transform in the adjoint representation of $SU(N)$ and $SU(3)$ or $SU(2)$ from the SM sector. This is an unusual case which cannot be embedded in string theory constructions (where messengers/open strings can only transform as bifundamentals, or symmetric antisymmetric, adjoint representations of a single non-abelian group), however, it could not be excluded in the framework of 4D QFTs.

\subsubsection{The action in the low-energy  regime}

The interaction terms are similar to \eqref{couplingsONLYMESONS}, \eqref{SMandMesonsMesino} where
\bea
&&\Phi_1\sim \Phi~~,~~~~~~~\Phi_2\sim X~~,~~~~~~~ \Psi_{1,2}\sim \Psi
\eea
since $\Psi=\Psi^c$ and the indexes $i,j$ for the fermions should be omitted.

\subsubsection{Seesaw mechanism with real mesinos from the messenger sector}

In comparison to the previous case \eqref{ActionMASS}, we have a single Majorana mesino $\Psi=\Psi^c$ and the mass generating terms are
\bea
S_{\text{mass}} =\int d^4x \bigg( g_{LH\Psi} {v_2} ~ \overline{{\n}}_L \Psi + M_M \Psi\Psi \bigg)~~~~~
\eea
with $v=g_{LH\Psi} {v_2}$ and $M_M\sim M$ the heavy messenger scale.
The mass matrix is given by
\bea
{\cal M} =
\left(\ba{cccc}
0& v \\
v& M_M
\ea \right)
\label{MassSeesaw_Majorana}\eea
with the standard eigenvalues
$v^2/M_M ,~M_M$.
The masses, mixings of light, and sterile neutrinos are given by the standard Majorana type I seesaw mechanism, and they have been evaluated and analyzed in numerous works in the past.

\section{Three neutrino model}\label{3 Neutrino Cases}

In this section, we will generalize our model to obtain, after the seesaw mechanism, three light neutrinos, and some heavy sterile neutrinos.
The most straightforward generalization of the model in section \ref{SOMEexamples} is to assume one fermionic and three different bosonic messengers.

\subsection*{Messengers in complex representations}

First, we consider messengers in complex representation that transform like
\bea
\ba{llllccccccccclcccc}
 & & & &SU(N) && SU(3) &&SU(2) &&~~~Y~~~& & ~~~& \\
& & \phi^i_{N,q}~:~~~& (&\Box_N&;& \cdot &,&\cdot &,&q&)&~~~~~~~~&\textrm{for $i=1,2,3$}\\
& & \psi_{N,q}~:~~~& (&\Box_N&;& \cdot &,&\cdot &,&q&)&
\ea
\label{messengersCHARGEDmodel2a}
\eea
and their charge-conjugates.
The mesons and mesinos out of the messengers above are singlets under the hidden and SM gauge groups.
\bea
\ba{llllllllllll}
\Phi^{ij} &\sim & \phi^i_{N,q} \phi^{j*}_{N ,q} ~&
&~~ \textrm{boson}\\
\Psi^i &\sim & \phi_{N,q}^{i*}\psi_{N ,q} ~&
&~~ \textrm{fermion}\\
\Psi^{ic} &\sim & \phi^i_{N,q} \psi_{N ,q}^c ~&
&~~ \textrm{fermion}\\
X &\sim & \psi_{N,q}\bar \psi_{N ,q} ~&
&~~ \textrm{boson}
\ea
\label{MesonsMesionosMODELmany}\eea
This model has ten mesons $\Phi^{ij},X$ and three mesinos $\Psi_i$. The generalization of the formulae \eqref{couplingsONLYMESONS},\eqref{SMandMesonsMesino} is strait forward. Focusing on the fermionic part we have
\bea
\mathcal{S}^{\text{eff}}_{\text{Meson}} =\int d^4x \bigg(\sum_{ij=1}^3 \Big( g_{ij} {v_2} ~ \overline{{\n}}_{iL} \Psi_i + g_{ij} {v_2} ~ \overline{\n}_{iL} \Psi_j^c
\Big)
+ M^{ij}_D \bar \Psi_i\Psi_j + M^{ij}_M \bar \Psi_i^c\Psi_j+ h.c.\bigg)~~~~~~
\label{Seesaw for many neutrinos}\eea
where the scale of the messengers is again $M$, however, the couplings in \eqref{Seesaw for many neutrinos} take different values.
The mass matrix is $9\times 9$,
\bea
{\cal M} =
\left(\ba{cccc}
0& v & v' \\
v^T& M_D & M_M \\
v'^T& M^T_M & M_D
\ea \right)
=
\left(\ba{ccc}
0& V_{3\times 6} \\
V_{6\times 3}^T& \tilde M_{6\times 6}
\ea \right)
\eea
where $V_{3\times 6}= (v_{3\times 3},v'_{3\times 3})$ terms after EW symmetry breaking and $\tilde M_{6\times 6}=\left(\ba{cccc}
M_D & M_M \\
M^T_M & M_D
\ea \right)$.
In order to diagonalize this mass matrix, we follow the procedure described in \cite{Dev:2012sg, Grimus:2000vj, Hettmansperger:2011bt} using a unitary matrix
\bea
{\cal V}=\left(\ba{cccc}
(\mathbb{1}_3 +\x^* \x^T)^{-1/2}& \x^* (\mathbb{1}_6 +\x^T \x^*)^{-1/2} \\
-\x^T (\mathbb{1}_3 +\x^* \x^T)^{-1/2}& (\mathbb{1}_6 +\x^T \x^*)^{-1/2}
\ea \right)
\eea
with an arbitrary $3\times 6$ matrix $\x$. The block diagonalization condition implies that
\bea
V-\x \tilde M - \x V^T \x^* = \mathbb{0}_{3\times 6}
\eea
which can be solved in terms of $V$, $\tilde M$. At leading order, we have $\x=V \tilde M^{-1}$, and the light neutrino mass matrix simplifies to
\bea
M_\n = - V \tilde M^{-1} V^T
\eea
triggering the seesaw mechanism since $\tilde M$ contains the heavy messenger scale.

Assuming that this model is realized within semi-realistic D-brane constructions, mesinos are charged under some anomalous $U(1)$ symmetries broken via the Green-Schwarz mechanism. Therefore, Majorana mass terms $M_M$ are much lower than the heavy messenger scale $M$ and consequently $M_D$ since $M_M\ll M_D\sim M$. This case falls in the models studied in \cite{Dev:2012sg}.

\subsection*{Messengers in real representations}

If the gauge group of the hidden sector is $SO(2N+1)$ or $Sp(N)$, we can consider a single fermionic and three copies of bosonic messengers transforming like
\bea
\ba{llllccccccccclcccc}
 & & & &SO(2N+1) && SU(3) &&SU(2) &&~~~Y~~~& & ~~~& \\
& & \phi^i~:~~~& (&V_{2N+1}&;& \cdot &,&\Box_2&,&0&)&~~~~~~~~&\textrm{for $i=1,2,3$}\\
& & \psi~:~~~& (&V_{2N+1}&;& \cdot &,&\Box_2&,&0&)&
\ea
\label{messengersREALmodel2}
\eea
and the real mesons/mesinos
\bea
\ba{llllllllllll}
\Phi^{ij} &\sim & \phi^i \phi^j ~&
&~~ \textrm{bosons}\\
\Psi^i &\sim & \phi^i\psi ~&
&~~ \textrm{fermions}\\
X &\sim & \bar \psi \psi ~&
&~~ \textrm{boson}
\ea
\label{MesonsMesionosMODELreal}\eea
with three Majorana mesinos $\Psi^i$. The mass matrix is a $6\times 6$ matrix with the form of \eqref{MassSeesaw_Majorana} where each block is a $3\times 3$ matrix. That case leads to the standard seesaw mechanism, which gives three light SM neutrinos and three heavy sterile neutrinos.

\vskip 2cm

\section*{Acknowledgements}\label{ACKNOWL}
\addcontentsline{toc}{section}{Acknowledgements}

\noindent We would like to thank Elias Niederwieser for participating in the early stages of this work.
We would also like to thank I. Antoniadis, P. Betzios, D. Consoli, Y. Mambrini, A. Pilaftsis for discussions.

\noindent E.K. was partially supported by an advanced ERC grant SM-GRAV, No 669288.  P.A. was supported by FWF Austrian Science Fund via the SAP P30531-N27.

\appendix
\renewcommand{\theequation}{\thesection.\arabic{equation}}
\addcontentsline{toc}{section}{Appendix\label{app}}
\section*{Appendix}

\section{Meson two-point functions} \label{a-2p}

In this appendix, we have determined the dependence of messenger mesons' effective action from the basic scales of the theory.

First, we focus on a single meson operator as follows
\be
S=S_{UV}+\int d^4 x \left[\Lambda^{4-\Delta_{\Lambda}}O_{\Lambda}+ \delta_{ij}\left(M^{4-\Delta}\Phi_{ij}\right)\right]
\label{a4a}\ee
Consider now the renormalized log of the partition function of this theory.
\be
\log Z(M,\Lambda)=V_4\Lambda^4f\left({M\over \Lambda}\right)
\label{a5}\ee
which has been written from general principles of scaling.
By the definition of the relevant couplings we must have\footnote{When the relevant operators have integer dimensions, scaling anomalies appear and the scalings in (\ref{a6}) are modified by logs. We will not consider this case as it does not change significantly the story, but we shall comment at the end.}
\be
 \log Z(M,\Lambda=0)=c_M~V_4~M^4\sp \log Z(M=0,\Lambda)=c_{\Lambda}V_4~\Lambda^4
\label{a6}\ee
where $c_{M,\Lambda}$ are dimensionless constants and depend on the parameters of the UV CFT, $V_4$ is the four-dimensional volume and we have normalized
\be
 \log Z(M=0,\Lambda=0)=0
 \label{a7}\ee
 This implies that
 \be
 \lim_{x\to 0}f(x)=c_{\Lambda}
 \sp
 \lim_{x\to\infty}f(x)\equiv {c_{M}}x^4
 \label{a8} \ee
 We now consider the relevant one-point functions in order to address the subleading terms in $f$.
We have from the definition
\be
{d\over d \Lambda^{4-\Delta_{\Lambda}}}\log Z=
{1\over (4-\Delta_{\Lambda})\Lambda^{3-\Delta_{\Lambda}}}{d\over d \Lambda}\log Z={V_4\Lambda^{\Delta_{\Lambda}}\over (4-\Delta_{\Lambda})}\left(4f-{M\over \Lambda}f'\right)=
\int d^4x \langle O_{\Lambda}\rangle =V_4 \langle O_{\Lambda}\rangle
 \label{a11}\ee

\be
{d\over d M^{4-\Delta}}\log Z=
{1\over (4-\Delta)M^{3-\Delta}}{d\over d M}\log Z={V_4M^{\Delta-3}\Lambda^{3}\over (4-\Delta)}\left(f'\right)=
\int d^4x \langle O\rangle =V_4 \langle O\rangle
 \label{a12}\ee

By taking $M\to 0$ in (\ref{a11}) and using (\ref{a8}) we obtain
\be
\lim_{M\to 0}\langle O_{\Lambda}\rangle={4c_{\Lambda}\over 4-\Delta_{\Lambda}}\Lambda^{\Delta_{\Lambda}}
 \label{a13}\ee
By taking $\Lambda\to 0$ in (\ref{a12}) and using (\ref{a8}) we obtain
\be
\lim_{\Lambda\to 0}\langle O\rangle={4c_{M}\over 4-\Delta}M^{\Delta}
 \label{a14}\ee
which are consistent with standard scaling.
Now we compute $\langle O\rangle$ at $M=0$ and we expect that
\be
\langle O\rangle=c_1\Lambda^{\Delta}
 \label{a15}\ee
This is consistent with  (\ref{a12}) if
\be
f(x)=c_{\Lambda}+c_1x^{4-\Delta}+o(x^{3-\Delta})
 \label{a16}\ee
 which improves (\ref{a8}).
Similarly, at $\Lambda=0$, we have
\be
\langle O_{\Lambda}\rangle=c_2M^{\Delta_{\Lambda}}
 \label{a17}\ee
which via (\ref{a11}) implies that near $x\to\infty$
\be
f(x)=x^4\left[c_{M}+c_2x^{\Delta_{\Lambda}-4}+\cdots\right]
 \label{a18}\ee
 which also improves (\ref{a8}).

We now consider two-point functions,
\be
\int d^4x\int d^4y\langle O_{\Lambda}(x)O(y)\rangle={d\over d M^{4-\Delta}}{d\over d \Lambda^{4-\Delta_{\Lambda}}}\log Z
={V_4 M^{\Delta-3}\Lambda^{\Delta_{\Lambda}-1}\over  (4-\Delta_{\Lambda}) (4-\Delta)}\left(3f'-xf''\right)\Big|_{x={M\over \Lambda}}
 \label{a19}\ee
Near $x\to 0$ we obtain
\be
\int d^4x \langle O_{\Lambda}(x)O(0)\rangle=c_1{\Delta\over 4-\Delta_{\Lambda}}\Lambda^{\Delta+\Delta_{\Lambda}-4}
 \label{a20}\ee
while near $x\to\infty$ we obtain
\be
\int d^4x \langle O_{\Lambda}(x)O(0)\rangle=c_2{\Delta_{\Lambda}\over 4-\Delta}M^{\Delta+\Delta_{\Lambda}-4}
 \label{a21}\ee

Let us denote
\be
G(x-y)\equiv \langle O(x) O(y)\rangle_c
 \label{a9}\ee
and its momentum space version $G(p)$ where the subscript $c$ stands for ``connected".
From (\ref{a4a}) we obtain that
\be
\left({\delta\over\delta M^{4-\Delta}} \right)^2\log Z=\int d^4x\int d^4 y \langle O(x) O(y)\rangle_c=V_4 G(p=0)
 \label{a10}\ee
 so that
 \be
V_4 G(p=0)= {1\over (4-\Delta)^2 M^{6-2\Delta}}\left[{d^2\over dM^2}+{\Delta -3\over M}{d\over dM}\right]\log Z
 \label{a22}\ee
from which we obtain
\be
G(p=0)={\Lambda^3~M^{2\Delta-7}\over (\Delta-4)^2}\left[xf''+(\Delta-3)f'\right]_{x={M\over \Lambda}}
 \label{a23}\ee
Taking some limits we have
\bi
\item $M\gg \Lambda$ In this case, using (\ref{a18}) we obtain
\be
G(p=0)\simeq {4c_M\Delta\over (4-\Delta)^2}M^{2\Delta-4}
 \label{a24}\ee

\item $M\ll \Lambda$ In this case, the result depends on a subleading term from these in (\ref{a16}). We expect
\be
G(p=0)=d_1 \Lambda^{2\Delta-4}+\cdots
 \label{a25}\ee
and therefore (\ref{a16}) is corrected as follows
 \be
f(x)=c_{\Lambda}+c_1x^{4-\Delta}+{d_1\over 2}x^{2(4-\Delta)}+o(x^{2(4-\Delta)})
 \label{a16a}\ee
\ei

We shall now consider the ${\cal O}(p^2)$ terms in the correlator.
For this, we must consider the regime where $p^2$ is smaller than at least one of the two mass scales. Otherwise, the $p$ dependence is determined to leading order in the UV CFT, and its leading behavior is $p^{\Delta-4}$ unless when $\Delta$ is integer when logs appear.
For $p^2\ll max(M^2,\Lambda^2)$ we write
\be
G(p)=G(p=0)\left[1+{p^2\over M^2g(M/\Lambda)}+{\cal O}(p^4)\right]
 \label{a26}\ee
 In the two limits we obtain
 \be
 \lim_{x\to \infty}g(x)=-{1\over \kappa}\sim {\rm constant}\sp \lim_{x\to 0}g(x)\sim x^2
 \label{a26a} \ee
 We have
\bi
\item $M\gg \Lambda$, $p^2\ll M^2$
 \be
 G(p)=G(0)\left[1-\kappa{p^2\over M^2}+\cdots\right]\sim M^{2\Delta-4}\left[1-\kappa{p^2\over M^2}+\cdots\right]
 \label{a27}\ee
\item $M\ll \Lambda$, $p^2\ll \Lambda^2$
 \be
 G(p)=G(0)\left[1-\kappa{p^2\over \Lambda^2}+\cdots\right]\sim \Lambda^{2\Delta_{\Lambda}-4}\left[1-\kappa{p^2\over \Lambda^2}+\cdots\right]
 \label{a28} \ee
However, for $X'$, the flavor non-singlet part of $X$, the estimate is different as they are pseudo-Goldstone bosons. In that case
\be
 G_{X'}(p)\sim \Lambda^{2\Delta_{\Lambda}-4}\left[1-\kappa{p^2\over M\Lambda}+\cdots\right]
 \label{a29} \ee
The propagating fields associated with the mesonic operators $\chi_{\Phi,\Psi,X}$ are the Legendre transforms, and therefore their quadratic action is given by the inverse of the associated two-point function.
\ei

\subsection{The effective action for mesons et mesinos}

We shall now include the mesinos.
 We have denote the associated spinor operators as $\Psi^{\a}_{i\bar j}$ in (\ref{a2}). In a weak coupling formulation, $\Psi^{\a}_{i\bar j}$ can be thought of as a colorless composite of a fermionic and a scalar quark.
We still have the action (\ref{a1}) and we have already shown that the two point functions of $X_{i\bar j},\Phi_{i\bar j}$ depend on $M,\Lambda$ as in
\be
\langle \Phi_{i\bar j}\Phi_{k\bar l}\rangle\sim \delta_{i,\bar l}\delta_{\bar j,k}M^{2\Delta_{\Phi}-4}\left[1-\kappa{p^2\over M^2}+\cdots\right]\sp M\gg \Lambda\sp p^2\ll M^2
 \label{a27a}\ee
The singlet piece $X\equiv \delta^{i\bar j}X_{i\bar j}$ has a similar two-point function to (\ref{a27a}).
The non-singlet piece $X'_{i\bar j}=X_{i\bar j}-{\delta_{i\bar j}\over N}X$ on the other hand has the following two-point function
\be
\langle X'_{i\bar j}X'_{k\bar l}\rangle\sim \left(\delta_{i,\bar l}\delta_{\bar j,k}-{\delta_{i\bar j}\delta_{k\bar l}\over N}\right)M^{2\Delta_{X}-4}\left[1-\kappa'{p^2\over M\Lambda}\cdots\right]\sp M\gg \Lambda\sp p^2\ll M^2
 \label{a27b}\ee
as in (\ref{a29}).
Finally, for the mesinos we obtain
\be
\langle \Psi^{\a}_{i\bar j}\Psi^{\b}_{k\bar l}\rangle\simeq
\delta_{i,\bar l}\delta_{\bar j,k}M^{2\Delta_{\Psi}-4}\left[1-\kappa''{\slashed{p}^{\a\b}\over M}+\cdots\right]\sp M\gg \Lambda\sp p^2\ll M^2
\label{a28aa}\ee

\subsection{From 2-point fucntions to kinetic terms}

Next, we want to derive the form of the action for the composite fields $\Phi$, $\Psi$, $X$.

\bi
\item The $\Phi$, $X$ (we denote both by a single field $\chi$ and $\Delta_\chi$ the dimension)

Returning to configuration space, we obtain for $M\gg \Lambda$
\bea
S_{\Phi,X}&=&\int d^4x \left[1-\kappa{\square\over M^2}\right]{\left(\chi(x)-{4c_M\over 4-\Delta_\chi}M^{\Delta_\chi}\right)^2\over M^{2\Delta_\chi-4}}
\label{SPhiX_1}\eea
Notice that the field $\chi(x)$ has a non vanishing vev $\langle \chi(x) \rangle = {4c_M\over 4-\Delta_\chi}M^{\Delta_\chi}$ which we subtract in \eqref{SPhiX_1}.
Therefore, we can define canonically normalized fields with zero vev as
\bea
{\cal X}\equiv
{\sqrt{\k}\over M^{\Delta_\chi-1}}\chi(x)-{4c_M\sqrt{\k}\over 4-\Delta}{M}
 \label{a31}\eea
so that
\be
S_{\Phi,X}=
 -\int d^4p \left(-{M^2\over \k }+{\square}\right){{\cal X}(x)^2}
 \label{a32}\ee
Adding non-derivative non-linear interactions we obtain
\be
S_{\Phi,X}=
 -\int d^4x \left(-{M^2\over \k}+{\square}\right) {\cal X}(x)^2
 +\sum_{n=3}^{\infty}g_n{M^{4-n}\over \k^{n\over 2}} {\cal X}(x)^n
 \label{a33}\ee

\item For the non-singlet mesons $X$ (which we denote by $X'$), the mass term $M^2$ should be replaced by $M\Lambda$. We have
\bea
S_{X'}&=&\int d^4x \left[1-\kappa'{\square\over M\Lambda}\right]{\left(X'(x)-{4c_M\over 4-\Delta_{X'}}M^{\Delta_{X'}}\right)^2\over M^{2\Delta_{X'}-4}}
 \label{a30}
\eea
We can define canonically normalized fields with zero vev as
\bea
\bar{\cal X}\equiv {\sqrt{\k'} \over M^{\Delta_{X'}-{3\over 2} }\Lambda^{1\over 2}}X' -{4c_M\sqrt{\k'}\over 4-\Delta_{X'}}\left({M^3\over \Lambda}\right)^{1\over 2}
\label{a34}
\eea
so that
adding non-derivative non-linear interactions we get
\bea
S_{X'}= -\int d^4p \left(-{M\Lambda\over \k'}+{\square}\right){{\cal X}'(x)^2}+\sum_{n=3}^{\infty}g_n{M^{4-n}\over (\k')^{n\over 2}} {\cal X}'(x)^n
\label{a36}\eea

\item The $\Psi$

Therefore, the action for the mesinos (kinetic term) would be\footnote{We omit delta functions of the two-point function since the quadratic action contains only the same fields.}
\bea
S_\Psi&=& \int d^4x M^{4-2\D_\Psi} \bar \Psi (1 - i \slashed \partial {\k'' \over M})\Psi \\
&=& -\int d^4x \k'' M^{3-2\D_\Psi} \bar \Psi ( i \slashed \partial -{M\over \k''})\Psi \\
&=& -\int d^4x \bar {\widetilde\Psi} ( i \slashed \partial-{M\over \k''} )\widetilde \Psi
 \label{a37}\eea
after the canonical normalization 
\bea
{\widetilde \Psi}= \sqrt {\k''} M^{{3\over 2}-\D_\Psi} \Psi \label{a38}
\eea
Adding non-derivative non-linear interactions
\bea
S_{\widetilde\Psi}&=& -\int d^4x \bar {\widetilde\Psi} ( i \slashed \partial-{M\over \k''} )\widetilde \Psi + \sum_{n=2}^\infty q_n M^{4-3n} (\bar {\widetilde\Psi} \widetilde \Psi)^n
\label{a39} \eea

\ei


\subsection{Scale for mixed terms}\label{Scale for mixed terms}

In this appendix, we will evaluate the scale for the mixed couplings between the mesons/mesinos and SM operators. For simplicity, we will focus on the simplest coupling between the messenger fields and the Higgs $H H^* \Phi$ and $H H^* \Phi^2$.

Since $\Psi\sim \phi \bar \phi$ we have to go one step back, and build diagrams including $H$, $\phi$ and $\bar \phi$.
In general, $H$ and $\phi$ are charged under $U(1)$ or $U(1)'$. Therefore, we can build diagrams where the associated gauge field is exchanged.
The couplings to be used are $A_\m (p^\m H)H^*$, $A^2 HH^*$, $A_\m (p^\m \phi) \bar \phi$ and $A^2 \phi \bar \phi$.
\bi
\item The $H H^* \Phi$ comes from $H H^* \phi \bar \phi$. The simplest diagram is at tree level where we have the exchange of a gauge field $A$ that belongs in the SM sector.
\bea
\scalebox{0.8}{
\begin{tikzpicture}[thick,baseline={-0.1cm}]
  \begin{feynman}[every blob={/tikz/fill=gray!30,/tikz/inner sep=2pt}]
    \vertex (i1) at (-1, 0.75) {\(H\)};
    \vertex (i2) at (-1,-0.75) {\(H^*\)};
    \vertex (f1) at (0,0);
    \vertex (f2) at (1,0);
    \vertex (e1) at (2, 0.25){\(\phi\)};
    \vertex (e2) at (2,-0.25){\(\bar \phi\)};
    \vertex (e6) at (2.60, 0){\(\bigg\} \Phi\)};
    \diagram* {
      (i1) --  (f1),
      (i2) --  (f1),
      (f1) -- [boson,edge label=\(A\)] (f2),
      (e1) --  (f2),
      (e2) --  (f2)}; \end{feynman}
\end{tikzpicture}}
\eea
Both couplings contain the momenta $\sim p$ of one of the incoming/outgoing scalars. The coupling is proportional to the
\bea
p^2 H H^* {1\over p^2}\phi \bar \phi ~~ \sim ~~ M H H^* \left(\frac{\phi \bar \phi}{M}\right)  ~~\sim ~~ M H H^* \Phi
\eea
where the $1/p^2$ comes from the propagator of the gauge field.


For higher contributions, a 1-loop diagram with the gauge field $A$ in the loop:
\bea
\scalebox{0.8}{
\begin{tikzpicture}[thick,baseline={-0.1cm}]
  \begin{feynman}[every blob={/tikz/fill=gray!30,/tikz/inner sep=2pt}]
    \vertex (i1) at (-1, 0.75) {\(H\)};
    \vertex (i2) at (-1,-0.75) {\(H^*\)};
    \vertex (f1) at (0,0);
    \vertex (f2) at (1,0);
    \vertex (e1) at (2, 0.25){\(\phi\)};
    \vertex (e2) at (2,-0.25){\(\bar \phi\)};
    \vertex (e6) at (2.60, 0){\(\bigg\} \Phi\)};
    \diagram* {
      (i1) --  (f1),
      (i2) --  (f1),
      (f1) -- [boson,half right] (f2),
      (f2) -- [boson,half right, edge label'=\(A\)] (f1),
      (e1) --  (f2),
      (e2) --  (f2)}; \end{feynman}
\end{tikzpicture}}
\eea
has degree of divergence zero, therefore the scale of the coupling is logarithmic, and after normalization we have
\bea
M^0 H H^* \phi \bar \phi ~~ \sim ~~ M H H^* \left(\frac{\phi \bar \phi}{M}\right)  ~~\sim ~~ M H H^* \Phi
\eea

\item Higher couplings like $HH^* \Phi^n$ with $n\ge2$.

The simplest diagram is a 1-loop diagram with the gauge field $A$ in the loop. For $n=2$ we have
\bea
\scalebox{0.8}{
\begin{tikzpicture}[thick,baseline={-0.1cm}]
  \begin{feynman}[every blob={/tikz/fill=gray!30,/tikz/inner sep=2pt}]
    \vertex (i1) at (-1, 0.75) {\(H\)};
    \vertex (i2) at (-1,-0.75) {\(H^*\)};
    \vertex (f1) at (0,0);
    \vertex (f2) at (2,0.75);
    \vertex (f3) at (2,-0.75);
    \vertex (e1) at (3.25, 1.50){\(\phi\)};
    \vertex (e2) at (3, 1.75){\(\bar \phi\)};
    \vertex (e3) at (3.25, -1.50){\(\bar \phi\)};
    \vertex (e4) at (3, -1.75){\(\phi\)};
    \vertex (e5) at (3.80, 1.625){\(\bigg\} \Phi\)};
    \vertex (e6) at (3.80, -1.625){\(\bigg\} \Phi\)};
    \diagram* {
      (i1) --  (f1),
      (i2) --  (f1),
      (f1) -- [boson, quarter left, edge label=\(A\)] (f2),
      (f2) -- [boson, quarter left] (f3),
      (f3) -- [boson, quarter left] (f1),
      (e1) --  (f2),
      (e2) --  (f2),
      (e3) --  (f3),
      (e4) --  (f3)}; \end{feynman}
\end{tikzpicture}}
\eea
and the degree of divergence is $2$. Therefore,
\bea
M^{-2} H H^* \phi \bar \phi \phi \bar \phi ~~ \sim ~~  H H^* \left(\frac{\phi \bar \phi}{M}\right)\left(\frac{\phi \bar \phi}{M}\right)  ~~\sim ~~  H H^* \Phi^2
\eea
I use as scale for the coupling $\Lambda$ (SM scale) since the particles in the loop are coming from the SM sector.

For generic $n$ we have superficial degree of divergence $2n-2$.
\bea
M^{-2n+2} H H^* (\phi \bar \phi)^n ~~ \sim ~~  M^{2-n} H H^* \left(\frac{\phi \bar \phi}{M}\right)^n  ~~\sim ~~ M^{2-n} H H^* \Phi^n
\eea
in agreement with \eqref{SMmesinosMIXED}.

\ei


\section{A string-inspired Standard Model}\label{D-brane SM}

In this section, we embed our configuration in a semi-realistic D-brane SM. In these realizations, SM particles are located on a set of D-brane stacks (minimum three stacks and, in most cases, four or more). Gauge fields are open strings with both end-points on the same stack of D-brane, while matter fields are stretched between different stacks.

Among the plethora of different realizations, we consider a bifundamental realization\footnote{This is the type of realizations that arise in orientifolds of string theory. They are also the most general realizations where all SM fields are bifundamentals of the gauge group. The gauge groups embedding of the hypercharge in such realizations were classified in \cite{ADKS}. } of the SM, charged under four abelian gauge fields with charge operators $Q_3,~Q_2,~Q_1,~Q_1'$. The hypercharge is given by the linear combination of these four gauge fields as
\bea
Y={1\over 6} Q_3 + {1\over 2} Q_1 - {1\over 2} Q_1'
\label{hyperchargeMadrid}
\eea
The SM fields are charged under these $U(1)$'s and the non-abelian gauge groups $SU(2)$ and $SU(3)$ as
\bea
\ba{llrrrrrrrrrrrrrrr}
&Q_L&({\bf 3},&{\bf 2},&1/6)&: &~~(&1,& -1 ,&0 ,&0&)\\
&u_R^c&(\bar{\bf 3},&{\bf 1},&-2/3)&: &~~ (&-1,& 0,& 0,& 1&) \\
&d_R^c&(\bar{\bf 3},&{\bf 1},&1/3)&: &~~ (&-1,& 0 ,&0 ,&-1&) \\
&L_L&({\bf 1},&{\bf 2},&-1/2)&: &~~(&0,& -1,& -1,& 0&) \\
&l_R^c&({\bf 1},&{\bf 1},& 1)&: &~~(&0,&0,& 1,& -1&) \\
&H_1&({\bf 1},&{\bf 2},&-1/2)&: &~~ (&0,& 1,& 0,& 1&) \\
&H_2&({\bf 1},&{\bf 2},& 1/2)&: &~~(&0,& 1,& 0,& -1&)
\ea
\label{MadridCharges}\eea
where we have expressed the whole spectrum in $L$-spinors (conjugate of $R$-handed are $L$-handed).
All SM Yukawa couplings are allowed for this spectrum, assuming they are present at the perturbative level.


The simplest model with emergent neutrinos contains only two messengers, one fermionic and one bosonic, which have the following charge assignments
\bea
\phi_{a\bar 1}~:~ (N; -1) &=& (\Box; 0,0,-1,0) ~~,~~~~~ \psi_{a1'}~:~ (N; +1') = (\Box; 0,0,0,1)
\label{messengersCHARGEDmodel1a}
\eea
and their charge-conjugates. The $\Box$ denotes that these messengers transform in the fundamental of the hidden $SU(N)$ gauge group.

The mesons and mesinos out of the messengers above \eqref{messengersCHARGEDmodel1a} are
\bea
\ba{lllrrrrrrrrrrrrrr}
\Phi_{1,\bar 1} &\sim & \phi_{a\bar 1} \phi^*_{a\bar 1} ~&~ (0;& 0,&0,&0,&0) ~&~~ \textrm{boson}\\
\Psi_{1,1'} &\sim & \phi_{a\bar 1}^*\psi_{a 1'} ~&~ (0;& 0,&0,&1,&1) ~&~~ \textrm{fermion}\\
\Psi_{1,1'}^c &\sim & \phi_{a\bar 1} \psi_{a 1'}^c ~&~ (0;& 0,&0,&-1,&-1) ~&~~ \textrm{fermion}\\
X_{1',\bar 1'} &\sim & \psi_{a1'}\bar \psi_{a 1'} ~&~ (0;& 0,&0,&0,&0) ~&~~ \textrm{boson}
\ea
\label{MesonsMesionosMODEL1string}\eea
Next, we provide the leading couplings of these mesons to various SM operators.

The $\Psi_{1,1'}^c$ mesino in \eqref{MesonsMesionosMODEL1} can couple to the $\bar L_LH_2^*$ coupling and therefore, it can play the role of the right-handed neutrino.


\end{document}